\documentclass[twocolumn,seceq]{jpsj2}

\title{Multipole Ordering and Fluctuations in $f$-Electron Systems}

\author{Katsunori \textsc{Kubo} and Takashi \textsc{Hotta}}

\inst{Advanced Science Research Center, Japan Atomic Energy Agency,
Tokai, Ibaraki 319-1195}

\recdate{\today}

\abst{
We investigate effects of multipole moments in $f$-electron systems
both from phenomenological and microscopic viewpoints.
First, we discuss significant effects of octupole moment
on the magnetic susceptibility in a paramagnetic phase.
It is found that even within mean-field approximation,
the magnetic susceptibility deviates from the Curie-Weiss law
due to interactions between dipole and octupole moments.
Next, we proceed to a microscopic theory for multipole ordering
on the basis of a $j$-$j$ coupling scheme.
After brief explanation of a method to derive multipole interactions
from the $f$-electron model, we discuss several multipole ordered phases
depending on lattice structure.
Finally, we show our new development of the microscopic approach
to the evaluation of multipole response functions.
We apply fluctuation exchange approximation to the $f$-electron model,
and evaluate multipole response functions.
}

\kword
{octupole moment, $j$-$j$ coupling scheme, Curie-Weiss law,
fluctuation exchange approximation}

\begin{document}
\maketitle

%%%%%%%%%%%%%%%%%%%%%%%%%%%%%%%%%%%%%%%%%%%%%%%%%%%%%%%%%%%%%%%%%%%%%%%%%%%%
%
% Sec.I Introduction
%
%%%%%%%%%%%%%%%%%%%%%%%%%%%%%%%%%%%%%%%%%%%%%%%%%%%%%%%%%%%%%%%%%%%%%%%%%%%%
\section{Introduction}

Novel magnetism and exotic superconductivity in rare-earth~\cite{orbital2001}
and actinide~\cite{Santini2} compounds have attracted much attention
in the research field of condensed matter physics.
In order to elucidate the mechanism of these phenomena,
it is quite important to describe $f$-electron states in a systematic way
by considering correctly the symmetry of $f$-electron orbitals.
On the other hand, we encounter difficulties to express
$f$-electron states by means of a simple combination of spin and orbital
degrees of freedom, as has been done for $d$-electron states,
since the spin-orbit interaction is strong for $f$ orbitals in general.
For such a case, it is rather convenient to exploit $multipole$ degrees
of freedom, including charge (rank-0), dipole (rank-1),
quadrupole (rank-2), and octupole (rank-3) components.
Fortunately, among such multipole moments,
dipole and quadrupole correspond
to ordinary spin and orbital degrees of freedom, respectively.
Thus, phenomena concerning dipole and/or quadrupole moments
are basically understood by analogy with $d$-electron physics.
In this sense, octupole is considered to be the first exotic
multipole moment in the solid state physics.

Although it has been very rare to observe phenomena related to
octupole moment in $d$-electron compounds, we have sometimes
found octupole-related phenomena in $f$-electron compounds.
For instance, in order to understand anomalous properties of CeB$_6$,
significant roles of octupole moment have been discussed actively.
In particular, apparent contradiction between neutron scattering
and NMR measurements on the so-called phase II of CeB$_6$ has been
clearly resolved by considering effects of magnetic-field induced
octupole moment on the hyperfine coupling.~\cite{Sakai:hf}
In recent years, furthermore, possibility of ordering of octupoles
themselves has been intensively discussed for
Ce$_x$La$_{1-x}$B$_6$~\cite{Kuramoto,Kusunose,Kubo,Kubo2,Mannix,Kusunose2}
and for
NpO$_2$.~\cite{Santini,Santini3,Paixao,Caciuffo,Lovesey,Kiss,Tokunaga,Sakai2}
In both materials, the crystalline electric field (CEF) ground states
are $\Gamma_8$ quartets,~\cite{Zirngiebl,Sato,Luthi,Fournier,Amoretti2}
and in such CEF ground states, there are dipole, quadrupole,
and octupole degrees of freedom.~\cite{Shiina}
In fact, several experimental facts have been reconciled
by assuming $\Gamma_{5u}$ antiferro-octupole ordering
in these materials.

In order to contribute to the further development of multipole physics,
there are two different ways for the direction of research.
One is to clarify the effect of octupole moment in a paramagnetic phase,
not in the ordered state, in a phenomenological level.
It is one of natural extensions of the research for mutipoles,
since the phenomenological theory for octupole ordering
has been developed so far.
Here it should be noted that octupole moments are magnetic,
that is, parity of octupoles is odd under time reversal.
Thus, in principle, octupoles would influence magnetic properties
even in the paramagnetic phase.
However, such effects have not been studied seriously.
We should make efforts to include such effects to analyze magnetic
properties of materials with active octupole degrees of freedom.

Another direction of the research is to develop a microscopic theory
for multipole-related phenomena.
The construction of such a microscopic theory was considered
to be difficult, since it has been believed that it is standard
to use the $LS$ coupling scheme to express the $f$-electron state.
However, we have recently developed a microscopic theory
for $f$-electron systems by exploiting a $j$-$j$ coupling
scheme.~\cite{Hotta,Kubo:NpO2,Kubo:jj}
The ground and low-energy excited states can be well described
by a microscopic model on the basis of the $j$-$j$ coupling scheme,
in contrast to the naive belief for the limitation
of the $j$-$j$ coupling scheme.
For instance, we have successfully explained the microscopic origin
of octupole ordering of NpO$_2$ even in a simple
$d$-electron-like approximation.~\cite{Kubo:NpO2,Kubo:jj}

Since we have shown that the microscopic model
on the basis of the $j$-$j$ coupling scheme is useful
to understand multipole ordering for $f$-electron systems,
it is natural to investigate further the effect of
{\it multipole fluctuations} in a microscopic level.
We believe that it is challenging and important development
of the microscopic theory for $f$-electron materials.
In fact, it has been actually proposed that
exotic superconductivity in PrOs$_4$Sb$_{12}$~\cite{Bauer}
may be induced by quadrupole fluctuations,~\cite{Miyake,Goto,Kuwahara}
since a quadrupole ordered phase appears
under a magnetic field.~\cite{Y_Aoki,Kohgi}
In order to discuss such a possibility from a microscopic view point,
it is necessary to include multipole fluctuations
in a microscopic theory for superconductivity.
In particular, we should classify complex fluctuations
in $f$-electron systems into multiple components.

In this paper, we exhibit our new development for multipole physics
including review of our previous work.
In \S~\ref{sec:chi}, first we briefly review our simple phenomenological theory
to understand the effect of octupole moment in a paramagnetic phase.~\cite{Kubo:chi}
In particular, we discuss effects of octupoles on the magnetic
susceptibility, to clarify how octupoles influence macroscopic
properties in the paramagnetic phase.
In \S~\ref{sec:jj}, we explain the recent development of our microscopic
theory for multipole phenomena.
After a short explanation of the model on the basis
of the $j$-$j$ coupling scheme, we briefly review the microscopic
theory for multipole ordering.~\cite{Kubo:NpO2,Kubo:jj}
In particular, we find the $\Gamma_{5u}$ longitudinal
antiferro-octupole ordering for an fcc lattice,
which has been proposed for NpO$_2$.
Then, we explain new development of our theory to treat
multipole fluctuations.
We present our framework for the microscopic approach to
evaluate multipole response functions with some calculated results.
Finally, in \S~\ref{sec:summary}, we summarize this paper.
Throughout this paper, we use such units as $\hbar$=$k_{\rm B}$=1.

%%%%%%%%%%%%%%%%%%%%%%%%%%%%%%%%%%%%%%%%%%%%%%%%%%%%%%%%%%%%%%%%%%%%%%
%
% Sec. 2   Phenomenological approach
%
%%%%%%%%%%%%%%%%%%%%%%%%%%%%%%%%%%%%%%%%%%%%%%%%%%%%%%%%%%%%%%%%%%%%%%
\section{Phenomenological Approach}
\label{sec:chi}

In order to clarify complex effects of multipole moment,
it is useful to consider first the problem in a phenomenological level.
In fact, there are lots of literatures treating multipole ordering
along this direction and the phase diagram including multipole
ordered state has been successfully explained by applying
a mean-field approximation to the Heisenberg-like model
for multipole moments.
In this section, we briefly review our recent result based on
such a phenomenological model, to clarify significant effect of
octupole moment on magnetic properties in the paramagnetic phase
within the mean-field approximation.\cite{Kubo:chi}

Let us here consider a cubic system for simplicity.
There is no essential difficulty to extend the theory to other cases.
In a cubic system, the octupole moment $\mib{T}$ with the same
symmetry $\Gamma_{4u}$ as the dipole moment $\mib{J}$
is an independent degree of freedom in a certain CEF ground state,
e.g., $\Gamma_8$ ground state.
In such a case, octupole and dipole moments interact each other.
In general, we should note that $\mib{T}$ is $not$ orthogonal to
$\mib{J}$.
Thus, we redefine $\mib{T}^{\prime}$, a linear combination 
of $\mib{T}$ and $\mib{J}$, which is orthogonal to $\mib{J}$,
i.e., $\text{tr}(J_{\alpha}T^{\prime}_{\alpha})$=0,
where $\alpha$ indicates a cartesian component and
$\text{tr}(\cdots)$ denotes the sum of the expectation values
over the CEF ground states.
%%%%%%
We normalize $\mib{T}^{\prime}$ so as to satisfy
$\text{tr}(J_{\alpha}J_{\alpha})
=\text{tr}(T^{\prime}_{\alpha}T^{\prime}_{\alpha})$.
%%%%%%

Then, a mean-field Hamiltonian under a magnetic field $\mib{H}$
along the $z$-direction up to $O(H)$ is given by
\begin{equation}
  \begin{split}
    \mathcal{H}_{\text{MF}}
    =&-HM_z-(g_J \mu_{\text{B}})^2
    [\lambda_{\text{d}} J_z \langle J_z \rangle
    +\lambda_{\text{o}} T^{\prime}_z \langle T^{\prime}_z \rangle \\
   &+\lambda_{\text{do}}(J_z \langle T^{\prime}_z \rangle
    +T^{\prime}_z \langle J_z \rangle)],
  \end{split}
  \label{eq:MF}
\end{equation}
where $M_z$ denotes magnetic moment,
$g_J$ is the Land\'{e}'s $g$-factor,
and $\mu_{\text{B}}$ is the Bohr magneton.
Phenomenological dipole-dipole, octupole-octupole, and dipole-octupole
interactions are expressed by $\lambda_{\rm d}$, $\lambda_{\rm o}$,
and $\lambda_{\rm do}$.
We have multiplied the factor $(g_J \mu_{\text{B}})^2$
to the interaction terms for convenience.
In general, there are other multipole interactions,~\cite{Sakai}
but the mean field from them cancel out within $O(H)$
due to the cubic symmetry. Thus, eq.~\eqref{eq:MF} does not change,
even if we consider other multipole interactions.

\begin{figure}
  \includegraphics[width=8cm]{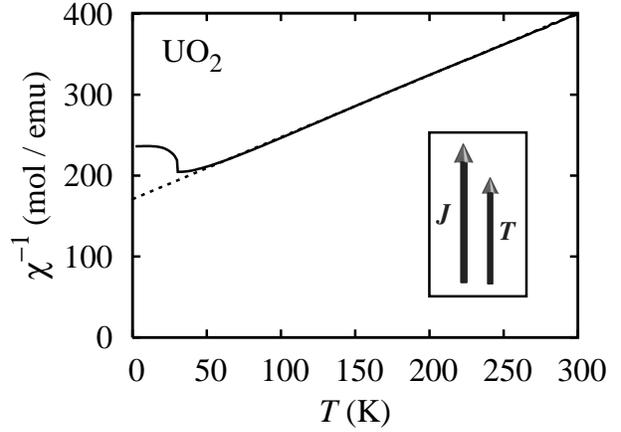}
  \caption{Temperature dependence of the inverse
    of the magnetic susceptibility $\chi$ of UO$_2$
    under $H$=10 kOe and $\mib{H} \parallel [100]$~\cite{Homma} (solid curve).
    The dotted curve represents the Curie-Weiss law with
    $\mu_{\text{eff}}=3.23\mu_{\text{B}}$
    [$C=\mu^2_{\text{eff}}/(3k_{\text{B}})$] and $\theta=-223$~K.
    The inset shows a schematic view of the dipole moment $\mib{J}$
    and octupole moment $\mib{T}$ in a $\Gamma_5$ CEF ground state.}
  \label{figure:UO2}
\end{figure}
\begin{figure}
  \includegraphics[width=8cm]{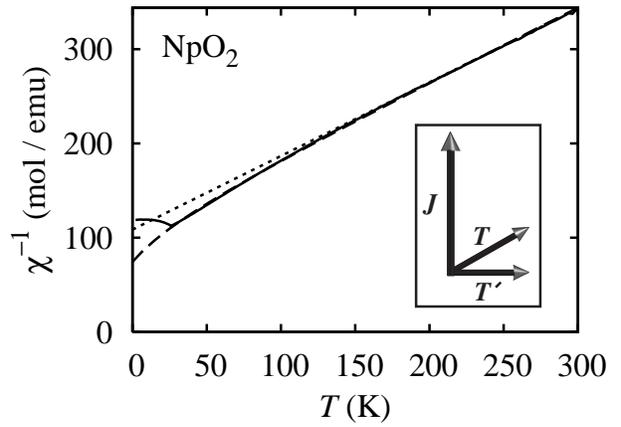}
  \caption{Temperature dependence of the inverse
    of the magnetic susceptibility $\chi$ of NpO$_2$
    under $H$=10 kOe and $\mib{H} \parallel [100]$~\cite{Aoki} (solid curve).
    The dashed curve represents the fit obtained with
    eq.~\eqref{modified_Curie-Weiss} with
    $\mu_{\text{eff}}=3.20 \mu_{\text{B}}$,
    $\theta=-146.1$~K, $\theta_{1}=-161.9$~K, and $\theta_2=-22.5$~K.
    The dotted curve represents the Curie-Weiss fit with
    $\mu_{\text{eff}}=3.20 \mu_{\text{B}}$ and $\theta=-139$~K.
    The inset shows a schematic view of the dipole moment $\mib{J}$,
    octupole moment $\mib{T}$,
    and $\mib{T}^{\prime}$ which is orthogonal to $\mib{J}$
    in a $\Gamma_8$ CEF ground state.}
  \label{figure:NpO2}
\end{figure}

Now we diagonalize the mean-field Hamiltonian by introducing
the linear combinations of $J_z$ and $T^{\prime}_z$:
\begin{align}
  M^{(1)}_z
  =-g_J \mu_{\text{B}} \cos \phi (\cos \phi J_z + \sin \phi T^{\prime}_z), \\
  M^{(2)}_z
  =-g_J \mu_{\text{B}} \sin \phi (\sin \phi J_z - \cos \phi T^{\prime}_z).
\end{align}
We have normalized $M^{(1)}_z$ and $M^{(2)}_z$
so that the sum of them is the magnetic moment $M_z$, i.e.,
$M^{(1)}_z+M^{(2)}_z=M_z=-g_J \mu_{\text{B}} J_z$.
By choosing the parameter $\phi$, we can diagonalize
the mean-field Hamiltonian as
\begin{equation}
  \begin{split}
    \mathcal{H}_{\text{MF}}
    =&-H M^{(1)}_z-\lambda_1 M^{(1)}_z \langle M^{(1)}_z \rangle \\
     &-H M^{(2)}_z-\lambda_2 M^{(2)}_z \langle M^{(2)}_z \rangle.
  \end{split}
\end{equation}
Then, the moment $M^{(1)}_z$ ($M^{(2)}_z$) follows the Curie-Weiss law
with the Curie constant $C_1=C \cos^2 \phi$ ($C_2=C \sin^2 \phi$)
and the Weiss temperature $\theta_1=C_1 \lambda_1$ ($\theta_2=C_2 \lambda_2$),
where $C$=$C_1+C_2$ is the Curie constant for a free ion.
The following relations for the Weiss temperatures
without the parameter $\phi$ may be useful:
$\theta_1+\theta_2$=$C(\lambda_{\rm d}+\lambda_{\rm o})$
and
$\theta_1\theta_2$=$C^2(\lambda_{\rm d}\lambda_{\rm o}-\lambda_{\rm do}^2)$.
Then, the magnetic susceptibility is given by
\begin{equation}
  \begin{split}
  \chi
  &=\frac{C_1}{T-\theta_1}+\frac{C_2}{T-\theta_2}\\
  &=C\frac{T+\theta-\theta_1-\theta_2}{(T-\theta_1)(T-\theta_2)},
  \end{split}
  \label{modified_Curie-Weiss}
\end{equation}
where $\theta$=$(C_1 \theta_1+C_2 \theta_2)/C$=$C\lambda_{\text{d}}$
is the Weiss temperature, i.e.,
the magnetic susceptibility follows the Curie-Weiss law
$\chi$=$C/(T-\theta)$ asymptotically at high temperatures.

As a typical application of the present theory,
we compare experimental results of
UO$_2$~\cite{Arrott,Homma} and NpO$_2$.~\cite{Ross,Erdos,Aoki}
In UO$_2$, the CEF ground state is a $\Gamma_5$ triplet.~\cite{Kern,Amoretti}
In this CEF ground state, the $\Gamma_{4u}$ octupole moment
is proportional to the dipole moment (see the inset of Fig.~\ref{figure:UO2}),
and thus, we expect Curie-Weiss behavior.
In Fig.~\ref{figure:UO2}, we show the temperature dependence of
the magnetic susceptibility of UO$_2$,
which follows the Curie-Weiss law very well
above the antiferromagnetic transition temperature 30~K at least up to 300~K.

In NpO$_2$, the CEF ground state is
a $\Gamma_8$ quartet.~\cite{Fournier,Amoretti2}
In this CEF ground state, the $\Gamma_{4u}$ octupole moment
is linearly independent of the dipole moment
(see the inset of Fig.~\ref{figure:NpO2}),
and the present theory is applicable.
Indeed, as shown in Fig.~\ref{figure:NpO2},
the magnetic susceptibility deviates from the Curie-Weiss law below 150~K.
This temperature is much higher than the transition temperature 25~K.
Moreover, this transition is probably a transition to an ordered state of
$\Gamma_{5u}$ octupole
moments,~\cite{Paixao,Caciuffo,Lovesey,Kiss,Tokunaga,Sakai2}
which do not influence
the magnetic susceptibility at least within the mean-field theory.
Thus, this deviation of the magnetic susceptibility from the Curie-Weiss law
is unlikely to stem from critical phenomena.
Rather, the present theory can naturally explain
this deviation, as shown in Fig.~\ref{figure:NpO2}.

%%%%%%%%%%%%%%%%%%%%%%%%%%%%%%%%%%%%%%%%%%%%%%%%%%%%%%%%%%%%%%%%%%%%%%%
%
%  Sec.III Microscopic theory
%
%%%%%%%%%%%%%%%%%%%%%%%%%%%%%%%%%%%%%%%%%%%%%%%%%%%%%%%%%%%%%%%%%%%%%%%
\section{Microscopic Approach}
\label{sec:jj}

In the previous section, we have briefly reviewed the phenomenological
approach to multipole physics, by focusing on the effect of octupole
moment on magnetic susceptibility in the paramagnetic phase.
In order to obtain a deeper insight into the phenomena concerning
multipole degrees of freedom, it is highly requested to proceed to
a microscopic theory beyond the phenomenological level.
For this purpose, we have recently developed the theory on the
basis of a $j$-$j$ coupling scheme.
After the explanation of the starting model,
we show our microscopic theory for multipole ordering
as well as recent development for the treatment of
multipole fluctuations.

%%%%%%%%%%%%%%%%%%%%%%%%%%%%%%%%%%%%%%%%%%%%%%%%%%%%%%%%%%%%%%%%%%%%%%%%
% 3-1 model
%%%%%%%%%%%%%%%%%%%%%%%%%%%%%%%%%%%%%%%%%%%%%%%%%%%%%%%%%%%%%%%%%%%%%%%%
\subsection{Model in a $j$-$j$ coupling scheme}

In order to include the effect of Coulomb interactions,
spin-orbit coupling, and CEF potential in $f$-electron systems,
the $LS$ coupling scheme has been frequently used.
However, it is not possible to apply
standard quantum-field theoretical technique
in the $LS$ coupling scheme, since Wick's theorem does not hold.
In order to overcome this difficulty, it has been proposed to
construct a microscopic model for $f$-electron systems
by exploiting the $j$-$j$ coupling scheme,~\cite{Hotta}
where we include first the spin-orbit coupling so as
to define the state labeled by the total angular momentum $j$.
For $f$ orbitals with angular momentum $\ell$=3,
we immediately obtain an octet with $j$=7/2(=3+1/2) and
a sextet with $j$=5/2(=3$-$1/2),
which are well separated by the spin-orbit interaction.
Since the spin-orbital coupling is, at least, in the order
of 0.1 eV for $f$ electrons, it would be enough to take into account
the $j$=5/2 sextet, when we investigate low-temperature properties
of $f$-electron compounds in the $j$-$j$ coupling scheme.

Then, the Hamiltonian is written as
\begin{equation}
  \begin{split}
  \mathcal{H}
  = & \sum_{\mib{r},\mib{a},\mu,\nu}
  t^{\mib{a}}_{\mu\nu}
  a^{\dagger}_{\mib{r} \mu}a_{\mib{r}+\mib{a} \nu}
  +\sum_{\mib{r},\mu,\nu}
  B_{\mu\nu} a_{\mib{r} \mu}^{\dag} a_{\mib{r} \nu} \\
  &+ (1/2) \sum_{\mib{r},\mu_1 \text{--} \mu_4}
  I_{\mu_1, \mu_2, \mu_3, \mu_4}
  a_{\mib{r} \mu_1}^{\dag} a_{\mib{r} \mu_2}^{\dag}
  a_{\mib{r} \mu_3} a_{\mib{r} \mu_4},
  \end{split}
  \label{Hjj}
\end{equation}
where $a_{\mib{r} \mu}$ is the annihilation operator
of the $f$ electron with the $z$-component $\mu$ of the total angular momentum
$j$=5/2 at site $\mib{r}$,
$\mib{a}$ denotes a vector connecting nearest-neighbor sites,
$t^{\mib{a}}_{\mu\nu}$ is the hopping integral between
electron for $\nu$-state at site $\mib{r}+\mib{a}$
and that for $\mu$-state at $\mib{r}$,
$B_{\mu\nu}$ is the CEF potential,
and the Coulomb integral $I$ in the $j$-$j$ coupling scheme
is expressed by using three Racah parameters,
$E_0$, $E_1$, and $E_2$.
In this study, we use a simple form for the hopping integrals, i.e.,
we consider the hopping integrals through the $(ff\sigma)$ bonding.

%%%%%%%%%%%%%%%%%%%%%%%%%%%%%%%%%%%%%%%%%%%%%%%%%%%%%%%%%%%%%%%%%%%%%%%%
% 3-2 ordering
%%%%%%%%%%%%%%%%%%%%%%%%%%%%%%%%%%%%%%%%%%%%%%%%%%%%%%%%%%%%%%%%%%%%%%%%
\subsection{Multipole ordering}
\label{sec:order}

Concerning multipole ordering,
we have already reported a microscopic approach
by considering correct $f$-electron symmetry.~\cite{Kubo:NpO2,Kubo:jj}
Thus, in this subsection, we briefly review
the method and obtained multipole ordered states.

Let us consider again a cubic system for simplicity.
In this case, it is convenient to express $f$-electron state
in terms of the eigenstates, $\Gamma_7$ and $\Gamma_8$,
in a cubic CEF potential.
Annihilation operators for the $\Gamma_8$ states are given by
\begin{subequations}
  \begin{align}
   f_{\mib{r} \alpha \uparrow}
   &= \sqrt{5/6} a_{\mib{r} 5/2}+\sqrt{1/6} a_{\mib{r} -3/2}, \\
   f_{\mib{r} \alpha \downarrow}
   &= \sqrt{5/6} a_{\mib{r} -5/2}+\sqrt{1/6} a_{\mib{r} 3/2},
  \end{align}
\end{subequations}
and
\begin{subequations}
  \begin{align}
   f_{\mib{r} \beta \uparrow} &= a_{\mib{r}  1/2},\\
   f_{\mib{r} \beta \downarrow} &= a_{\mib{r} -1/2},
  \end{align}
\end{subequations}
where we introduce pseudo-spin index $\sigma$(=$\uparrow$ and $\downarrow$)
to distinguish the states in each Kramers doublet,
while orbital index $\tau$ distinguishes degenerate Kramers doublets.
Annihilation operators for the $\Gamma_7$ states
are given by
\begin{subequations}
\begin{align}
  f_{\mib{r} \gamma \uparrow}
  &= \sqrt{1/6} a_{\mib{r} 5/2}-\sqrt{5/6} a_{\mib{r} -3/2},\\
  f_{\mib{r} \gamma \downarrow}
  &= \sqrt{1/6} a_{\mib{r} -5/2}-\sqrt{5/6} a_{\mib{r} 3/2}.
\end{align}
\end{subequations}

Now we consider the reduction of the Hamiltonian by suppressing
$\Gamma_7$ orbital, in particular, for actinide dioxides.
The reasons are as follows:
First, we note that the $\Gamma_7$ wave-function extends
along the [111] direction and other equivalent directions.
Since in actinide dioxides, oxygen anions locate in these directions,
the $\Gamma_7$ state should be energetically penalized.
Thus, it is expected to be higher than the $\Gamma_8$ level.
Second, when we accommodate two, three, and four electrons
in the $\Gamma_8$ orbitals, the ground states are
$\Gamma_5$, $\Gamma^{(2)}_8$, and $\Gamma_1$, respectively.
In fact, these states have been found as the CEF ground states in
UO$_2$,~\cite{Kern,Amoretti}
NpO$_2$,~\cite{Fournier,Amoretti2}
and  PuO$_2$,~\cite{Kern2,Kern3} respectively.
Finally, the CEF excitation energy in PuO$_2$ is as large as 123 meV.
It is expected to be the same order of magnitude
as the splitting between the $\Gamma_8$ and $\Gamma_7$ levels,
and thus we can ignore the $\Gamma_7$ states at low temperatures.

After transforming the Hamiltonian eq.~\eqref{Hjj}
to that in the $\Gamma_7$-$\Gamma_8$ basis,
we simply discard the terms concerning $\Gamma_7$ orbital.
Then, we obtain the $\Gamma_8$-orbital Hubbard model as
\begin{equation}
  \begin{split}
    \mathcal{H}_8
    =& \sum_{\mib{r},\mib{a},\tau,\sigma,\tau^{\prime},\sigma^{\prime}}
    {\tilde t}^{\mib{a}}_{\tau \sigma; \tau^{\prime} \sigma^{\prime}}
    f^{\dagger}_{\mib{r} \tau \sigma}
    f_{\mib{r}+\mib{a} \tau^{\prime} \sigma^{\prime}}
    \\
    &+U \sum_{\mib{r},\tau}
    n_{\mib{r} \tau \uparrow} n_{\mib{r} \tau \downarrow}
    +U^{\prime} \sum_{\mib{r}}
    n_{\mib{r} \alpha} n_{\mib{r} \beta}
    \\
    &+ J \sum_{\mib{r},\sigma,\sigma^{\prime}}
    f^{\dagger}_{\mib{r} \alpha \sigma}
    f^{\dagger}_{\mib{r} \beta \sigma^{\prime}}
    f_{\mib{r} \alpha \sigma^{\prime}}
    f_{\mib{r} \beta \sigma}
    \\
    &+ J^{\prime}\sum_{\mib{r},\tau \ne \tau^{\prime}}
    f^{\dagger}_{\mib{r} \tau \uparrow}
    f^{\dagger}_{\mib{r} \tau \downarrow}
    f_{\mib{r} \tau^{\prime} \downarrow}
    f_{\mib{r} \tau^{\prime} \uparrow},
  \end{split}
  \label{Gamma8_model}
\end{equation}
where
${\tilde t}^{\mib{\mu}}_{\tau \sigma; \tau^{\prime} \sigma^{\prime}}$
is the hopping integral of an electron with
$(\tau^{\prime}, \sigma^{\prime})$ at site $\mib{r}$+$\mib{a}$
to the $(\tau, \sigma)$ state at $\mib{r}$,
$n_{\mib{r} \tau \sigma}$
=$f^{\dagger}_{\mib{r} \tau \sigma} f_{\mib{r} \tau \sigma}$,
and
$n_{\mib{r} \tau}=\sum_{\sigma} n_{\mib{r} \tau \sigma}$.
The coupling constants $U$, $U^{\prime}$, $J$, and $J^{\prime}$
denote the intra-orbital, inter-orbital, exchange,
and pair-hopping interactions, respectively.
These are expressed in terms of Racah parameters, and we obtain
the relation $U$=$U'+J+J'$.

In order to estimate interactions among multipoles
from the above $\Gamma_8$-orbital model,
we apply the second-order perturbation theory with respect
to the hopping integrals of $\Gamma_8$ electrons,
in a manner to estimate the superexchange interactions
in $d$-electron systems.
Here we note that the hopping integrals are depend on the pseudo-spin
$\sigma$, orbital $\tau$, and the direction of the hopping $\mib{a}$
due to the $f$-electron symmetry.
Thus, the hopping integrals depend on the lattice structures,
and we obtain different multipole interactions for different lattices.

The obtained multipole interaction models have been further
analyzed by using the mean-field theory.
For a simple cubic lattice, we find a $\Gamma_{3g}$ antiferro-quadrupole
ordering at a finite temperature, and as we lower the temperature further,
we find another transition to a ferromagnetic state.
For a bcc lattice, we find a $\Gamma_{2u}$ antiferro-octupole transition,
and a ferromagnetic transition follows it.
For an fcc lattice, we analyze the multipole interaction model carefully,
since the fcc lattice has geometrical frustration.
Thus, we first evaluate multipole correlation functions numerically,
and determine relevant interactions to the ground state.
Then, we apply mean-field theory to a simplified model including
only these relevant interactions, and find a $\Gamma_{5u}$
longitudinal antiferro-octupole ordering.
This $\Gamma_{5u}$ ordering has been proposed for the ordered state
of NpO$_2$, in which Np ions forms an fcc lattice
and the CEF ground state is a $\Gamma_8$ quartet.

%%%%%%%%%%%%%%%%%%%%%%%%%%%%%%%%%%%%%%%%%%%%%%%%%%%%%%%%%%%%%%%%%%%%%%%%
% 3-3 fluctuations
%%%%%%%%%%%%%%%%%%%%%%%%%%%%%%%%%%%%%%%%%%%%%%%%%%%%%%%%%%%%%%%%%%%%%%%%
\subsection{Multipole fluctuations}
\label{sec:FLEX}

In the previous subsection, we have reviewed our microscopic theory
for multipole ordering.
It is considered to be natural to attempt to extend our theory
to treat multipole fluctuations.
In order to clarify multipole fluctuations from a microscopic view point,
again we resort to a method which has been widely used in
the $d$-electron research field.
Namely, we apply the fluctuation exchange (FLEX) approximation
to the model based on the $j$-$j$ coupling scheme.

Although there is no essential difficulty to consider
a three-dimensional lattice, here we consider a two-dimensional
square lattice for simplicity.
By keeping both $\Gamma_7$ and $\Gamma_8$ orbitals,
we rewrite the model Hamiltonian \eqref{Hjj} as
\begin{equation}
  \begin{split}
  \mathcal{H}
  =& \sum_{\mib{r},\mib{a},\tau,\sigma,\tau^{\prime},\sigma^{\prime}}
  t^{\mib{a}}_{\tau \sigma; \tau^{\prime} \sigma^{\prime}}
  f^{\dagger}_{\mib{r} \tau \sigma}
  f_{\mib{r}+\mib{a} \tau^{\prime} \sigma^{\prime}}
  \\
  &+U \sum_{\mib{r},\tau}
  n_{\mib{r} \tau \uparrow} n_{\mib{r} \tau \downarrow}
  +U^{\prime}/2 \sum_{\mib{r},\tau \ne \tau^{\prime}}
  n_{\mib{r} \tau} n_{\mib{r} \tau^{\prime}}
  \\
  &+\Delta \sum_{\mib{r}} n_{\mib{r} \beta}
  +\Delta^{\prime} \sum_{\mib{r}} n_{\mib{r} \gamma},
  \end{split}
  \label{jj_model}
\end{equation}
where $U$=$U^{\prime}$=$E_0$.
Note that for simplicity, we consider the situation in which
one of Racah parameters $E_0$ is finite and
the other Racah parameters are zero,
since the interactions between them are complicated in general.
We believe that characteristics of multipole fluctuations will
be grasped even in this simplified model,
as long as we treat the $f$-electron symmetry correctly.
In eq.~\eqref{jj_model}, we have introduced CEF parameters
$\Delta$ and $\Delta^{\prime}$, where
$\Delta$ denotes the splitting energy between $\Gamma_8$ orbitals,
while $\Delta^{\prime}$ indicates the splitting between
$\Gamma_8^{\alpha}$ and $\Gamma_7$ orbitals.
Note also that $\Delta$ is introduced to include partly
the effect of a tetragonal CEF potential.

On a square lattice, the hopping integrals of $f$ electrons
through $(ff\sigma)$ bonding do not depend on the pseudo-spin
$\sigma$, and they are given by
\begin{equation}
  t^{(a,0)}=
  \begin{pmatrix}
    1 & -1/\sqrt{3} & 0 \\
   -1/\sqrt{3} & 1/3 & 0 \\
    0 & 0 & 0
  \end {pmatrix}
  t,
\end{equation}
and
\begin{equation}
  t^{(0,a)}=
  \begin{pmatrix}
    1 & +1/\sqrt{3} & 0 \\
    +1/\sqrt{3} & 1/3 & 0 \\
    0 & 0 & 0
  \end {pmatrix}
  t,
\end{equation}
where $t=9(ff\sigma)/28$ and $a$ is the lattice constant.
In this subsection, the energy unit is $t$ and $a$ is taken as unity.

In the present model, since the pseudo-spin $\sigma$ is
a conserved quantity, the Green's function does not depend on $\sigma$.
Then, the Green's function is represented by $3 \times 3$ matrix.
The Dyson-Gorkov equation in a matrix form is given by
\begin{equation}
  G(k)=G^{(0)}(k)+G^{(0)}(k)\Sigma(k)G(k),
  \label{Dyson-Gorkov}
\end{equation}
where $G(k)$ is the Green's function and
$G^{(0)}(k)$ is the non-interacting Green's function.
Here we introduced the abbreviation $k$=$(\mib{k},\text{i} \epsilon_n)$,
where $\mib{k}$ is momentum and
$\epsilon_n$=$(2n+1) \pi T$ is the Matsubara frequency for fermions
with an integer $n$ and a temperature $T$.
The self-energy $\Sigma(k)$ is given by
\begin{equation}
  \Sigma_{\tau_1 \tau_2}(k)
  =\frac{T}{N}\sum_{q \tau^{\prime}_1 \tau^{\prime}_2}
  G_{\tau^{\prime}_1 \tau^{\prime}_2}(k-q)
  V_{\tau_1 \tau^{\prime}_1; \tau_2 \tau^{\prime}_2}(q),
\end{equation}
where $N$ is the number of lattice sites,
$q$=$(\mib{q}, \text{i} \omega_m)$,
and $\omega_m=2m \pi T$ is the Matsubara frequency for bosons.
The fluctuation exchange interaction $V(q)$ is given by
\begin{equation}
  V(q)=\frac{3}{2}V^s(q)+\frac{1}{2}V^c(q),
\end{equation}
where
\begin{align}
  V^s(q)&=U^s \chi^s(q) U^s-\frac{1}{2}U^s \chi^{(0)}(q) U^s+U^s,\\
  V^c(q)&=U^c \chi^c(q) U^c-\frac{1}{2}U^c \chi^{(0)}(q) U^c-U^c.
\end{align}
The matrices $U^s$ and $U^c$ are given by
$U^s_{\tau \tau; \tau \tau}=U^c_{\tau \tau; \tau \tau}=U$,
$U^s_{\tau \tau^{\prime}; \tau \tau^{\prime}}
=-U^c_{\tau \tau^{\prime}; \tau \tau^{\prime}}=U^{\prime}$,
and $U^c_{\tau \tau; \tau^{\prime} \tau^{\prime}}=2U^{\prime}$,
where $\tau \ne \tau^{\prime}$,
and the other matrix elements are zero.
The susceptibilities $\chi^s(q)$ and $\chi^c(q)$ are given by
\begin{align}
  \chi^s(q)&=\chi^{(0)}(q)[1-U^s \chi^{(0)}(q)]^{-1},\\
  \chi^c(q)&=\chi^{(0)}(q)[1+U^c \chi^{(0)}(q)]^{-1},
\end{align}
where
\begin{equation}
  \chi^{(0)}_{\tau_1 \tau_2; \tau_3 \tau_4}(q)
  =-\frac{T}{N}\sum_{k}
  G_{\tau_1 \tau_3}(k+q)
  G_{\tau_4 \tau_2}(k).
\end{equation}
We solve the Dyson-Gorkov equation \eqref{Dyson-Gorkov} numerically.

When we obtain the self-consistent solution,
we can also evaluate the multipole response functions.
The multipole response functions are given by
\begin{equation}
  \chi^{\Gamma_{\gamma}}(\mib{q},\text{i} \omega_n)
  \!=\!
  \sum_{\mib{r}} \int^{\beta}_0 \text{d}\tau
  \text{e}^{-\text{i} \mib{q} \cdot \mib{r}
  +\text{i} \omega_n \tau}
  \langle \text{T}_{\tau}
  O^{\Gamma_\gamma}_{\mib{r}}(\tau)
  O^{\Gamma_\gamma}_{\mib{0}}(0)
  \rangle,
\end{equation}
where $\mib{0}$ denotes the origin and $\beta=1/T$.
The Heisenberg representation for the operator
$O^{\Gamma_\gamma}_{\mib{r}}$ is defined as
\begin{equation}
  O^{\Gamma_\gamma}_{\mib{r}}(\tau)
  =\text{e}^{\beta (\mathcal{H}-\mu \hat{N})}
  O^{\Gamma_\gamma}_{\mib{r}}
  \text{e}^{-\beta (\mathcal{H}-\mu \hat{N})},
\end{equation}
where $\mu$ is the chemical potential and
$\hat{N}=\sum_{\mib{r},\tau} n_{\mib{r} \tau}$
is the number operator of electrons.
The multipole operator at site $\mib{r}$ is given by
\begin{equation}
  O^{\Gamma_\gamma}_{\mib{r}}
  =\sum_{\tau_1 \tau_2 \sigma_1 \sigma_2}
  f^{\dagger}_{\mib{r} \tau_1 \sigma_1}
  O^{\Gamma_\gamma}_{\tau_1 \sigma_1; \tau_2 \sigma_2}
  f_{\mib{r} \tau_2 \sigma_2}.
\end{equation}
The $6 \times 6$ matrix $O^{\Gamma_\gamma}$ is obtained by using
the operator equivalent method for each multipole operator.
We normalize $O^{\Gamma_\gamma}$ so that
the sum of squares of eigenvalues is unity, for convenience.
It is useful to introduce the $3 \times 3$ matrices
$\mib{O}^{s \Gamma_\gamma}$ and $O^{c \Gamma_\gamma}$
which describe spin and charge parts of the matrix $O^{\Gamma_\gamma}$,
respectively:
\begin{align}
  \mib{O}^{s \Gamma_\gamma}_{\tau_1 \tau_2}
  &=\frac{1}{\sqrt{2}}\sum_{\sigma_1 \sigma_2}
  O^{\Gamma_\gamma}_{\tau_1 \sigma_1; \tau_2 \sigma_2}
  \mib{\sigma}_{\sigma_2 \sigma_1},\\
  O^{c \Gamma_\gamma}_{\tau_1 \tau_2}
  &=\frac{1}{\sqrt{2}}\sum_{\sigma_1 \sigma_2}
  O^{\Gamma_\gamma}_{\tau_1 \sigma_1; \tau_2 \sigma_2}
  \delta_{\sigma_2 \sigma_1},
\end{align}
where $\mib{\sigma}$ are the Pauli matrices.
In the FLEX approximation without the vertex corrections,
the multipole response functions are given by
\begin{equation}
  \begin{split}
    \chi^{\Gamma_{\gamma}}(q)
    =\sum_{\tau_1 \tau_2 \tau_3 \tau_4}
    \Bigl[
    &\chi^s_{\tau_1 \tau_2; \tau_3 \tau_4}(q)
    \mib{O}^{s \Gamma_\gamma}_{\tau_2 \tau_1} \cdot
    \mib{O}^{s \Gamma_\gamma}_{\tau_3 \tau_4} \\
    +&\chi^c_{\tau_1 \tau_2; \tau_3 \tau_4}(q)
    O^{c \Gamma_\gamma}_{\tau_2 \tau_1}
    O^{c \Gamma_\gamma}_{\tau_3 \tau_4} \Bigr].
  \end{split}
\end{equation}

\begin{figure}
  \includegraphics[width=8cm]{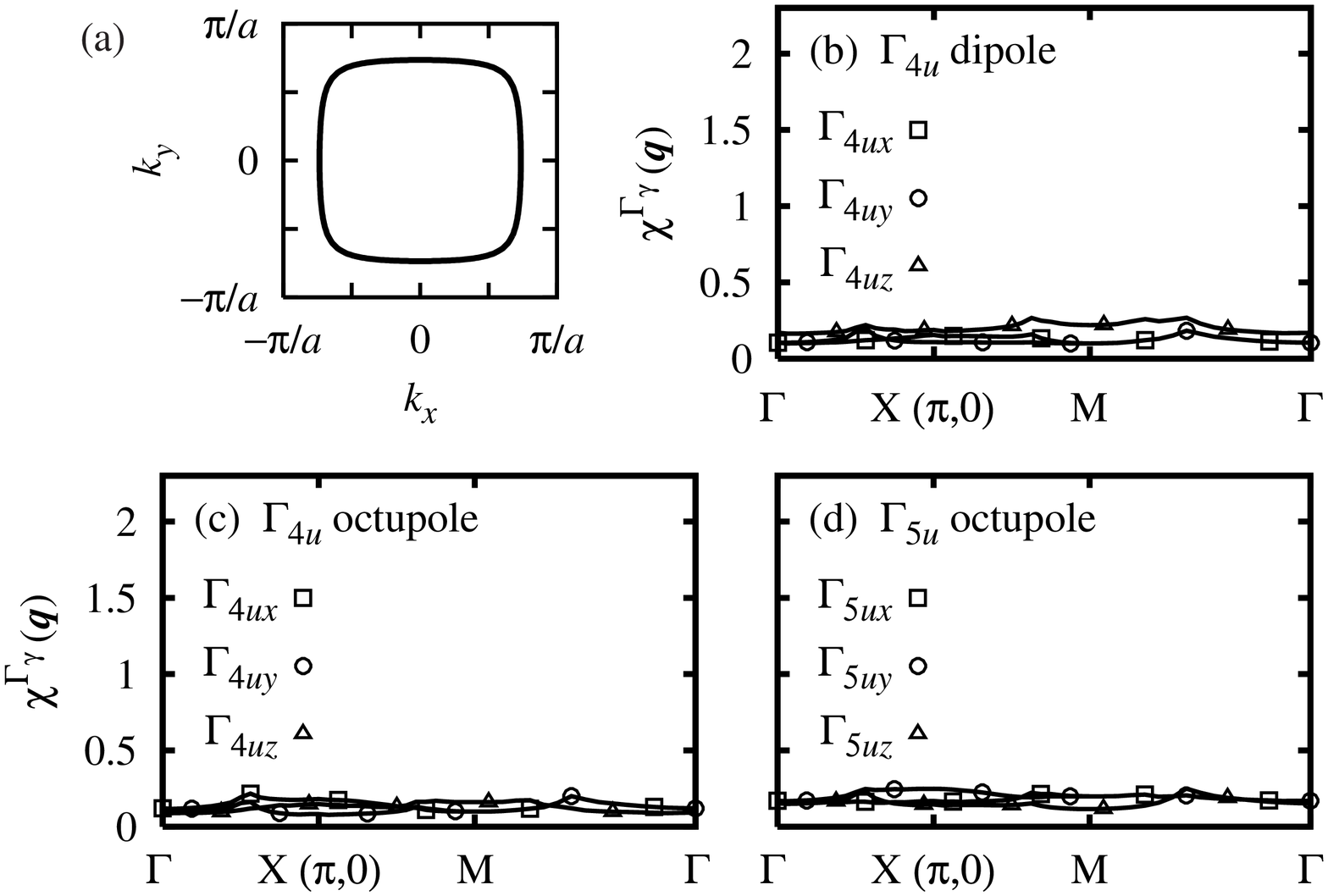}
%  {L5-N_tau10-N_matsu10-T_02-N1-U4-Up4-delta0-deltap0_2.eps}
  \caption{
    Fermi surface and multipole response functions for $\Delta=0$.
    (a) Fermi surface for $U=U^{\prime}=0$.
    (b) $\Gamma_{4u}$ dipole response functions.
    (c) $\Gamma_{4u}$ octupole response functions.
    (d) $\Gamma_{5u}$ octupole response functions.
  }
  \label{figure:delta0}
\end{figure}

\begin{figure}
  \includegraphics[width=8cm]{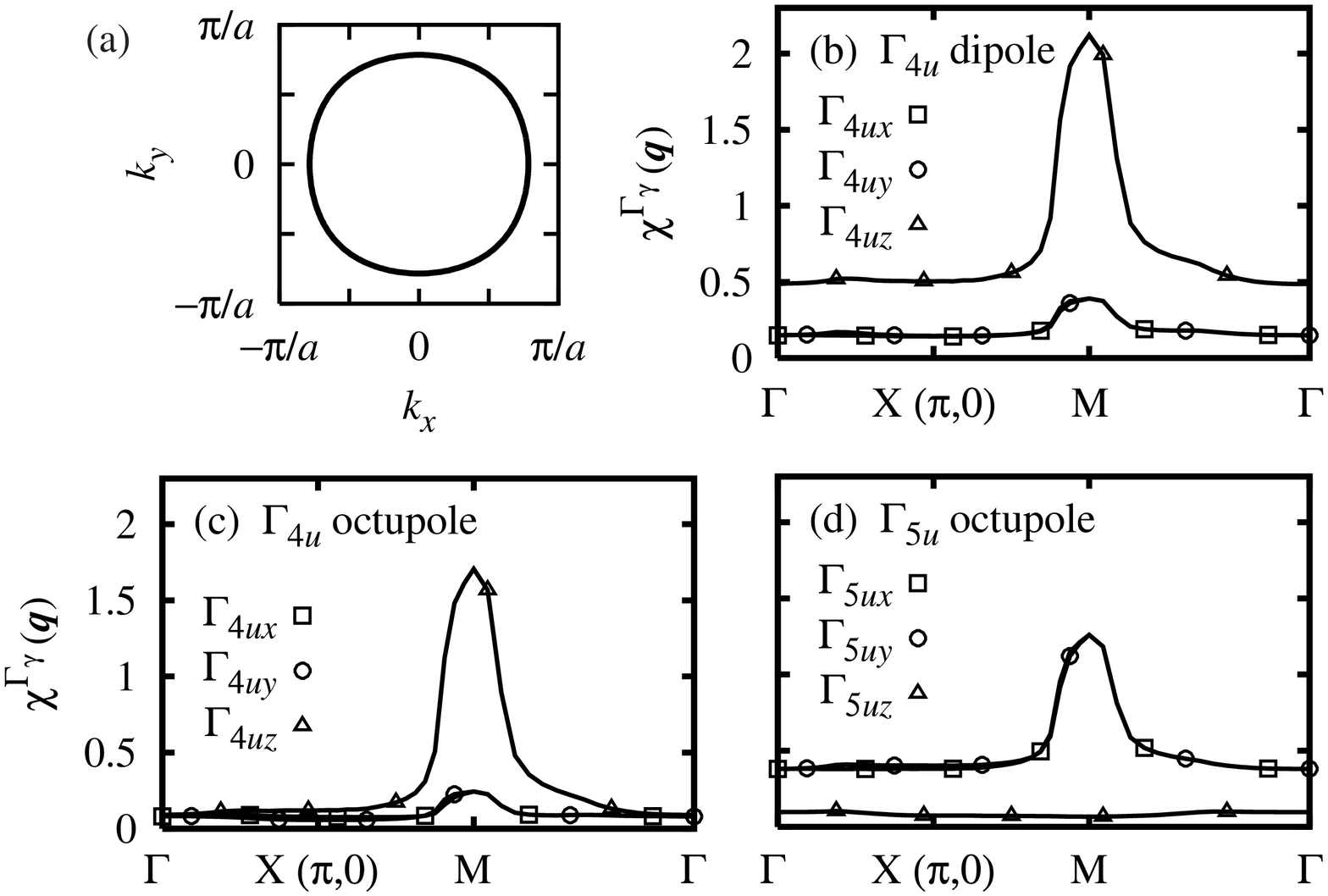}
%  {L5-N_tau10-N_matsu10-T_02-N1-U4-Up4-delta2-deltap0_2.eps}
  \caption{
    Fermi surface and multipole response functions for $\Delta=2t$.
    (a) Fermi surface for $U=U^{\prime}=0$.
    (b) $\Gamma_{4u}$ dipole response functions.
    (c) $\Gamma_{4u}$ octupole response functions.
    (d) $\Gamma_{5u}$ octupole response functions.
  }
  \label{figure:delta2}
\end{figure}

In the following, we show results for a $32 \times 32$ lattice,
at $U$=$U^{\prime}$=$4t$, $T$=$0.02t$, and $\Delta^{\prime}$=0.
The number of $f$ electrons per site is one.
In the calculation, we use 1024 Matsubara frequencies.
In Figs.~\ref{figure:delta0} and \ref{figure:delta2},
we show the multipole response functions for $\Delta$=0 and
$\Delta$=$2t$, respectively.
Even for $\Delta^{\prime}$=0,
the number $n_{\mib{r} \gamma}$ of electrons in the $\Gamma_7$ orbital
is almost zero for these model parameters,
since the $\Gamma_7$ orbital in the present model on a square lattice
is localized and the system cannot gain kinetic energy
by increasing the number of electrons in the $\Gamma_7$ orbital.
Thus, for these parameter sets, the present model is virtually
equivalent to the $\Gamma_8$ model,
which has been already studied by
Takimoto \textit{et al.}~\cite{Takimoto}
The antiferromagnetic fluctuations
pointed out by Takimoto \textit{et al.} can be classified
into multipole fluctuations in $f$-electron systems as shown in
Figs.~\ref{figure:delta0} and \ref{figure:delta2}.

For $\Delta$=0, there is no significant structures in the multipole response
functions, as shown in Figs.~\ref{figure:delta0}~(b)--(d).
By changing the CEF parameter $\Delta$, the Fermi surface changes
[see Figs.~\ref{figure:delta0}~(a) and \ref{figure:delta2}~(a)],
and the fluctuations around the M point [$\mib{q}=(\pi,\pi)$]
become significant.
Then, the multipole response functions for the $\Gamma_{4u}$ dipole,
$\Gamma_{4u}$ octupole, and $\Gamma_{5u}$ octupole moments
enhance around the M point.
On the other hand, the other multipole response functions do not enhance
significantly.
The $\Gamma_{4u}$ dipole and octupole components around at M point
are naturally considered to compose antiferromagnetic fluctuations,
but the enhancement of $\Gamma_{5u}$ octupole fluctuation
is quite peculiar.
It seems to be interesting to seek for anomalous component of
Cooper paring induced by $\Gamma_{5u}$ octupole fluctuation.
This is one of future problems.

\section{Summary}
\label{sec:summary}

In this paper, we have discussed three topics on multipoles
in $f$-electron systems.
First, we have explained effects of octupole moments
on the magnetic susceptibility in the paramagnetic phase
of a cubic system.
When the $\Gamma_{4u}$ octupole moment is independent of the dipole
moment as in the $\Gamma_8$ CEF ground state, there are interactions
between the $\Gamma_{4u}$ octupole and dipole moments in general.
Then, the magnetic susceptibility deviates from the Curie-Weiss law
even within the mean-field theory.
By considering such effects of octupoles,
we can naturally understand the difference between the temperature dependence
of the magnetic susceptibilities of UO$_2$ and NpO$_2$.

Second, we have derived multipole interactions from the microscopic model
based on the $j$-$j$ coupling scheme.
Then, we have found several multipole ordered states,
depending on the type of lattice structure.
In particular, for an fcc lattice, we find
the $\Gamma_{5u}$ longitudinal triple-$\mib{q}$ antiferro-octupole
ordering, which has been proposed for NpO$_2$ phenomenologically.

Finally, we have evaluated multipole response functions
by applying the FLEX approximations to the $j$=5/2 model
including both the $\Gamma_8$ and $\Gamma_7$ orbitals.
In $f$-electron systems, the complex fluctuations,
which are composed of spin, charge, and orbital degrees of freedom,
can be classified into multipole fluctuations.
We will study such multipole fluctuations
by changing the CEF parameters, dimensionality, and Coulomb
interactions, in order to clarify superconductivity
relating to the multipole fluctuations in the near future.

\section*{Acknowledgements}

We are grateful to T. Takimoto for useful comments and discussions
on the FLEX approximation.
We also thank D. Aoki and Y. Homma for providing us very recent
experimental results of the magnetic susceptibilities
in NpO$_2$ and UO$_2$, respectively, before publication.
We are also grateful to R. H. Heffner, S. Kambe, N. Metoki,
A. Nakamura, H. Onishi, Y. Tokunaga, R. E. Walstedt, and H. Yasuoka
for fruitful discussions.
One of the authors (K. K.) is supported by the REIMEI Research Resources
of Japan Atomic Energy Research Institute.
The other author (T. H.) is supported by a Grants-in-Aid
for Scientific Research in Priority Area ``Skutterudites''
under the contract No.~16037217 from the Ministry of
Education, Culture, Sports, Science, and Technology of Japan.
T. H. is also supported by a Grant-in-Aid for
Scientific Research (C)(2) under the contract No.~50211496
from Japan Society for the Promotion of Science.

%%%%%%%%%%%%%%%%%%%%%%%%%%%%%%%%%%%%%%%%%%%%%%%%%%%%%%%%%%%%%%%%%%%%%%%
%
% References
%
%%%%%%%%%%%%%%%%%%%%%%%%%%%%%%%%%%%%%%%%%%%%%%%%%%%%%%%%%%%%%%%%%%%%%%%

\end{document}